\documentclass[10pt,twocolumn,letterpaper]{article}

\usepackage{cvpr}
\usepackage{times}
\usepackage{graphicx}
\usepackage{amsmath}
\usepackage{amssymb}

\usepackage{url}
\usepackage{color}
\usepackage{enumitem}
\usepackage{multirow}
\usepackage{booktabs}
\usepackage{authblk}

\makeatletter
\renewcommand\AB@affilsepx{ --- \protect\Affilfont}
\renewcommand\@author{
  \def\rlap##1{##1}
  \vspace{-1cm}
  \url{http://www.shapenet.org}\\[0.2cm]
  \AB@authlist\\[\affilsep]
  \AB@affillist\\[0.1cm]
  Authors listed alphabetically
}
\makeatother

\usepackage[pagebackref=true,breaklinks=true,colorlinks,bookmarks=false]{hyperref}

\usepackage{cleveref}

\cvprfinalcopy 

\newcommand{\comment}[1]{}

\newcommand{\mypara}{\vspace*{-10pt}\paragraph}
\newcommand{\denselist}{\itemsep 0pt\parsep=0pt\partopsep 0pt\vspace{-\topsep}}
\newcommand\contactAuthorMark{\footnotemark[\arabic{footnote}]} 

\begin{document}

\title{ShapeNet: An Information-Rich 3D Model Repository}

\author[1]{Angel X. Chang}
\author[2]{Thomas Funkhouser}
\author[1]{Leonidas Guibas}
\author[1]{Pat Hanrahan}
\author[3]{Qixing Huang}
\author[3]{Zimo Li}
\author[1]{Silvio Savarese}
\author[1]{Manolis Savva\thanks{Contact authors: \texttt{\{msavva,haosu\}@cs.stanford.edu}}}
\author[2]{Shuran Song}
\author[1]{Hao Su\protect\contactAuthorMark}
\author[2]{Jianxiong Xiao}
\author[1]{Li Yi}
\author[2]{Fisher Yu}
\affil[1]{Stanford University}
\affil[2]{Princeton University}
\affil[3]{Toyota Technological Institute at Chicago}

\maketitle

\begin{abstract}
We present \emph{ShapeNet}: a richly-annotated, large-scale repository of shapes represented by 3D CAD models of objects.  ShapeNet contains 3D models from a multitude of semantic categories and organizes them under the WordNet taxonomy.  It is a collection of datasets providing many semantic annotations for each 3D model such as consistent rigid alignments, parts and bilateral symmetry planes, physical sizes, keywords, as well as other planned annotations.  Annotations are made available through a public web-based interface to enable data visualization of object attributes, promote data-driven geometric analysis, and provide a large-scale quantitative benchmark for research in computer graphics and vision. At the time of this technical report, ShapeNet has indexed more than 3,000,000 models, 220,000 models out of which are classified into 3,135 categories (WordNet synsets). In this report we describe the ShapeNet effort as a whole, provide details for all currently available datasets, and summarize future plans.
\end{abstract}

\section{Introduction}
Recent technological developments have led to an explosion in the amount of 3D data that we can generate and store.  Repositories of 3D CAD models are expanding continuously, predominantly through aggregation of 3D content on the web.  RGB-D sensors and other technology for scanning and reconstruction are providing increasingly higher fidelity geometric representations of objects and real environments that can eventually become CAD-quality models.

At the same time, there are many open research problems due to fundamental challenges in using 3D content.  Computing segmentations of 3D shapes, and establishing correspondences between them are two basic problems in geometric shape analysis.  Recognition of shapes from partial scans is a research goal shared by computer graphics and vision.  Scene understanding from 2D images is a grand challenge in vision that has recently benefited tremendously from 3D CAD models~\cite{Song:2014:slidingwindow,wu:2014:3d}. Navigation of autonomous robots and planning of grasping manipulations are two large areas in robotics that benefit from an understanding of 3D shapes.  At the root of all these research problems lies the need for attaching semantics to representations of 3D shapes, and doing so at large scale.

Recently, data-driven methods from the machine learning community have been exploited by researchers in vision and NLP (natural language processing). ``Big data'' in the visual and textual domains has led to tremendous progress towards associating semantics with content in both fields.  Mirroring this pattern, recent work in computer graphics has also applied similar approaches to specific problems in the synthesis of new shape variations~\cite{kalogerakis:2012:probabilistic} and new arrangements of shapes~\cite{fisher:2012:example}.  However, a critical bottleneck facing the adoption of data-driven methods for 3D content is the lack of large-scale, curated datasets of 3D models that are available to the community.

Motivated by the far-reaching impact of dataset efforts such as the Penn Treebank~\cite{marcus:1993:penntreebank}, WordNet~\cite{miller:1995:wordnet} and ImageNet~\cite{deng:2009:imagenet}, which collectively have tens of thousands of citations, we propose establishing \emph{ShapeNet}: a large-scale 3D model dataset.  Making a comprehensive, semantically enriched shape dataset available to the community can have immense impact, enabling many avenues of future research.

In constructing ShapeNet we aim to fulfill several goals:
\begin{itemize}\denselist
 \item Collect and centralize 3D model datasets, helping to organize effort in the research community.
 \item Support data-driven methods requiring 3D model data.
 \item Enable evaluation and comparison of algorithms for fundamental tasks involving geometry (e.g., segmentation, alignment, correspondence).
 \item Serve as a knowledge base for representing real-world objects and their semantics.
\end{itemize}

These goals imply several desiderata for ShapeNet:
\begin{itemize}\denselist
 \item Broad and deep coverage of objects observed in the real world, with thousands of object categories and millions of total instances.
 \item Categorization scheme connected to other modalities of knowledge such as 2D images and language.
 \item Annotation of salient physical attributes on models, such as canonical orientations, planes of symmetry, and part decompositions.
 \item Web-based interfaces for searching, viewing and retrieving models in the dataset through several modalities: textual keywords, taxonomy traversal, image and shape similarity search.
\end{itemize}

Achieving these goals and providing the resulting dataset to the community will enable many advances and applications in computer graphics and vision.

In this report, we first situate ShapeNet, explaining the overall goals of the effort and the types of data it is intended to contain, as well as motivating the long-term vision and infrastructural design decisions (\Cref{sec:shapenet}).  We then describe the acquisition and validation of annotations collected so far (\Cref{sec:acquisition}), summarize the current state of all available ShapeNet datasets, and provide basic statistics on the collected annotations (\Cref{sec:statistics}).  We end with a discussion of ShapeNet's future trajectory and connect it with several research directions (\Cref{sec:conclusion}).


\section{Background and Related Work}
There has been substantial growth in the number of of 3D models available online over the last decade, with repositories like the Trimble 3D Warehouse providing millions of 3D polygonal models covering thousands of object and scene categories.  Yet, there are few collections of 3D models that provide useful organization and annotations.  Meaningful textual descriptions are rarely provided for individual models, and online repositories are usually either unorganized or grouped into gross categories (e.g., furniture, architecture, etc.~\cite{GAMMA}).  As a result, they have been poorly utilized in research and applications.

There have been previous efforts to build organized collections of 3D models (e.g.,~\cite{AIM@SHAPE,GAMMA}). However, they have provided quite small datasets, covered only a small number of semantic categories, and included few structural and semantic annotations. Most of these previous collections have been developed for evaluating shape retrieval and classification algorithms. For example, datasets are created annually for the Shape Retrieval Contest (SHREC) that commonly contains sets of models organized in object categories. However, those datasets are very small --- the most recent SHREC iteration in 2014~\cite{li:2014:shrec} contains a ``large'' dataset with around 9,000 models consisting of models from a variety of sources organized into 171 categories (Table \ref{fig:shrec2014datasets}).

The Princeton Shape Benchmark is probably the most well-known and frequently used 3D shape collection to date (with over 1000 citations)~\cite{shilane:2004:princeton}.  It contains around 1,800 3D models grouped into 90 categories, but has no annotations beyond category labels.  Other commonly-used datasets contain segmentations~\cite{Chen:2009:segmentationbenchmark}, correspondences~\cite{Kim:2012:fuzzycorres, Kim:2013:partbasedtemplates}, hierarchies~\cite{Liu:2014:creating}, symmetries~\cite{Kim:2010:mobius}, salient features~\cite{Chen:2012:schelling}, semantic segmentations and labels~\cite{Xiao:2013:sun3d}, alignments of 3D models with images~\cite{xiang:2014:pascal3d}, semantic ontologies~\cite{AIM@SHAPE}, and other functional annotations --- but again only for small size datasets.  For example, the Benchmark for 3D Mesh Segmentation contains just 380 models in 19 object classes~\cite{Chen:2009:segmentationbenchmark}.

In contrast, there has been a flurry of activity on collecting, organizing, and labeling large datasets in computer vision and related fields.  For example, ImageNet \cite{deng:2009:imagenet} provides a set of 14M images organized into 20K categories associated with ``synsets'' of WordNet \cite{miller:1995:wordnet}.  LabelMe provides segmentations and label annotations of hundreds of thousands of objects in tens of thousands of images \cite{russell:2009:labelme}.  The SUN dataset provides 3M annotations of objects in 4K categories appearing in 131K images of 900 types of scenes. Recent work demonstrated the benefit of a large dataset of 120K 3D CAD models in training a convolutional neural network for object recognition and next-best view prediction in RGB-D data~\cite{wu:2014:3d}. Large datasets such as this and others (e.g.,~\cite{krause:2013:finegrain, liebelt:2010:multi}) have revitalized data-driven algorithms for recognition, detection, and editing of images, which have revolutionized computer vision.

Similarly, large collections of annotated 3D data have had great influence on progress in other disciplines.  For example, the Protein Data Bank \cite{berman:2000:pdb} provides a database with 100K protein 3D structures, each labeled with its source and links to structural and functional annotations \cite{laskowski:1997:pdbsum}.  This database is a common repository of {\em all} 3D protein structures solved to date and provides a shared infrastructure for the collection and transfer of knowledge about each entry. It has accelerated the development of data-driven algorithms, facilitated the creation of benchmarks, and linked researchers and industry from around the world. We aim to provide a similar resource for 3D models of everyday objects.

\begin{table*}
\centering{\small
\begin{tabular}{lcccc}
Benchmarks   &          Types           &      \# models       &       \# classes        & Avg \# models per class \\ \midrule
SHREC14LSGTB &         Generic          &        8,987         &           171           &           53            \\ \midrule
PSB          &         Generic          & 907+907 (train+test) &   90+92 (train+test)    &   10+10 (train+test)    \\
SHREC12GTB   &         Generic          &         1200         &           60            &           20            \\
TSB          &         Generic          &        10,000        &           352           &           28            \\
CCCC         &         Generic          &         473          &           55            &            9            \\
WMB          & Watertight (articulated) &         400          &           20            &           20            \\
MSB          &       Articulated        &         457          &           19            &           24            \\
BAB          &       Architecture       &         2257         & 183+180 (function+form) &  12+13 (function+form)  \\
ESB          &           CAD            &         867          &           45            &           19
\end{tabular}
}
\caption{\small{Source datasets from SHREC 2014: \emph{Princeton Shape Benchmark (PSB)}~\cite{shilane:2004:princeton}, \emph{SHREC 2012 generic Shape Benchmark (SHREC12GTB)}~\cite{li:2012:shrec}, \emph{Toyohashi Shape Benchmark (TSB)}~\cite{tatsuma:2012:large}, \emph{Konstanz 3D Model Benchmark (CCCC)}~\cite{vranic:2004:3d}, \emph{Watertight Model Benchmark (WMB)}~\cite{veltkamp:2007:shrec}, \emph{McGill 3D Shape Benchmark (MSB)}~\cite{zhang:2005:retrieving}, \emph{Bonn Architecture Benchmark (BAB)}~\cite{wessel:2009:3d}, \emph{Purdue Engineering Shape Benchmark (ESB)}~\cite{jayanti:2006:developing}.}}
\label{fig:shrec2014datasets}
\end{table*}

\section{ShapeNet: An Information-Rich 3D Model Repository}
\label{sec:shapenet}
ShapeNet is a large, information-rich repository of 3D models. It contains models spanning a multitude of semantic categories. Unlike previous 3D model repositories, it provides extensive sets of annotations for every model and links between models in the repository and other multimedia data outside the repository.

\begin{figure}
\includegraphics[width=\columnwidth]{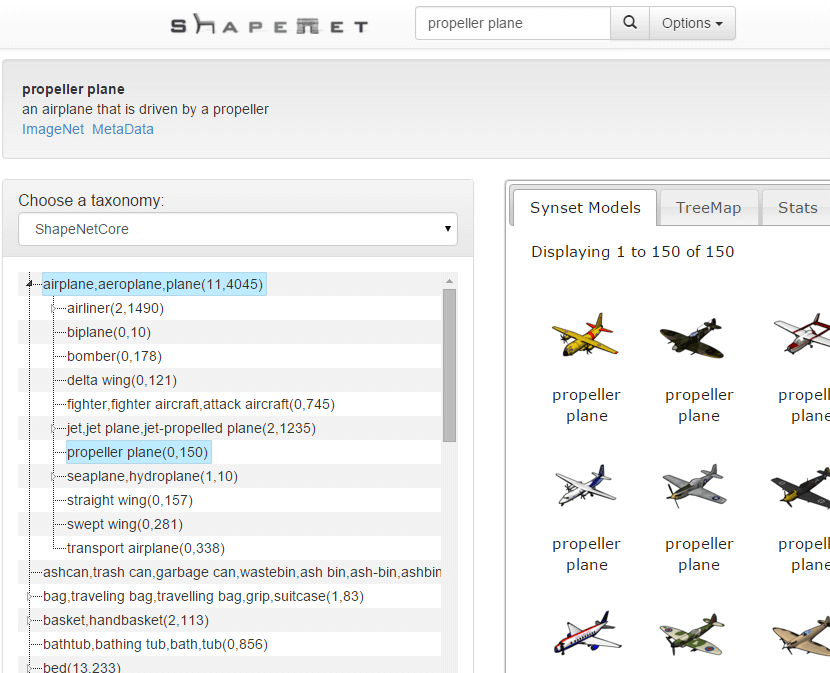}
\caption{Screenshot of the online ShapeNet taxonomy view, organizing contained 3D models under WordNet synsets.}
\label{fig:taxonomy-ui}
\end{figure}

Like ImageNet, ShapeNet provides a view of the contained data in a hierarchical categorization according to WordNet synsets (Figure~\ref{fig:taxonomy-ui}). Unlike other model repositories, ShapeNet also provides a rich set of annotations for each shape and correspondences between shapes. The annotations include geometric attributes such as upright and front orientation vectors, parts and keypoints, shape symmetries (reflection plane, other rotational symmetries), and scale of object in real world units. These attributes provide valuable resources for processing, understanding and visualizing 3D shapes in a way that is aware of the semantics of the shape.

We have currently collected approximately 3 million shapes from online 3D model repositories, and categorized 300 thousand of them against the WordNet taxonomy.  We have also annotated a subset of these models with shape properties such as upright and front orientations, symmetries, and hierarchical part decompositions. We are continuing the process of expanding the annotated set of models and also collecting new models from new data sources.

In the following sections, we discuss how 3D models are collected for ShapeNet, what annotations will be added, how those annotations will be generated, how annotations will be updated as the dataset evolves over time, and what tools will be provided for the community to search, browse, and utilize existing data, as well as contribute new data.

\subsection{Data Collection}

The raw 3D model data for ShapeNet comes from public online repositories or existing research datasets. ShapeNet is intended to be an evolving repository with regular updates as more and more 3D models become available, as more people contribute annotations, and as the data captured with new 3D sensors become prevalent.

We have collected 3D polygonal models from two popular public repositories: Trimble 3D Warehouse\footnote{https://3dwarehouse.sketchup.com/} and Yobi3D\footnote{https://yobi3d.com}. The Trimble 3D Warehouse contains 2.4M user-designed 3D models and scenes.  Yobi3D contains 350K additional models collected from a wide range of other online repositories. Together, they provide a diverse set of shapes from a broad set of object and scene categories --- e.g., many organic shape categories (e.g., humans and mammals), which are rare in Warehouse3D, are plentiful in Yobi3D.  For more detailed statistics on the currently available ShapeNet models refer to \Cref{sec:statistics}.

Though the tools developed for this project will be general-purpose, we intend to include only 3D models of objects encountered by people in the everyday world.  That is, it will not include CAD mechanical parts, molecular structures, or other domain-specific objects.  However, we will include scenes (e.g., office), objects (e.g., laptop computer), and parts of objects (e.g., keyboard).  Models are organized under WordNet~\cite{miller:1995:wordnet} noun ``synsets'' (synonym sets).  WordNet provides a broad and deep taxonomy with over 80K distinct synsets representing distinct noun concepts arranged as a DAG network of hyponym relationships (e.g., ``canary'' is a hyponym of ``bird'').  This taxonomy has been used by ImageNet to describe categories of objects at multiple scales~\cite{deng:2009:imagenet}.  Though we first use WordNet due to its popularity, the ShapeNet UI is designed to allow multiple views into the collection of shapes that it contains, including different taxonomy views and faceted navigation.

\subsection{Annotation Types}
\label{sec:annotation-types}

\begin{figure*}
  \centering
  \includegraphics[width=\textwidth]{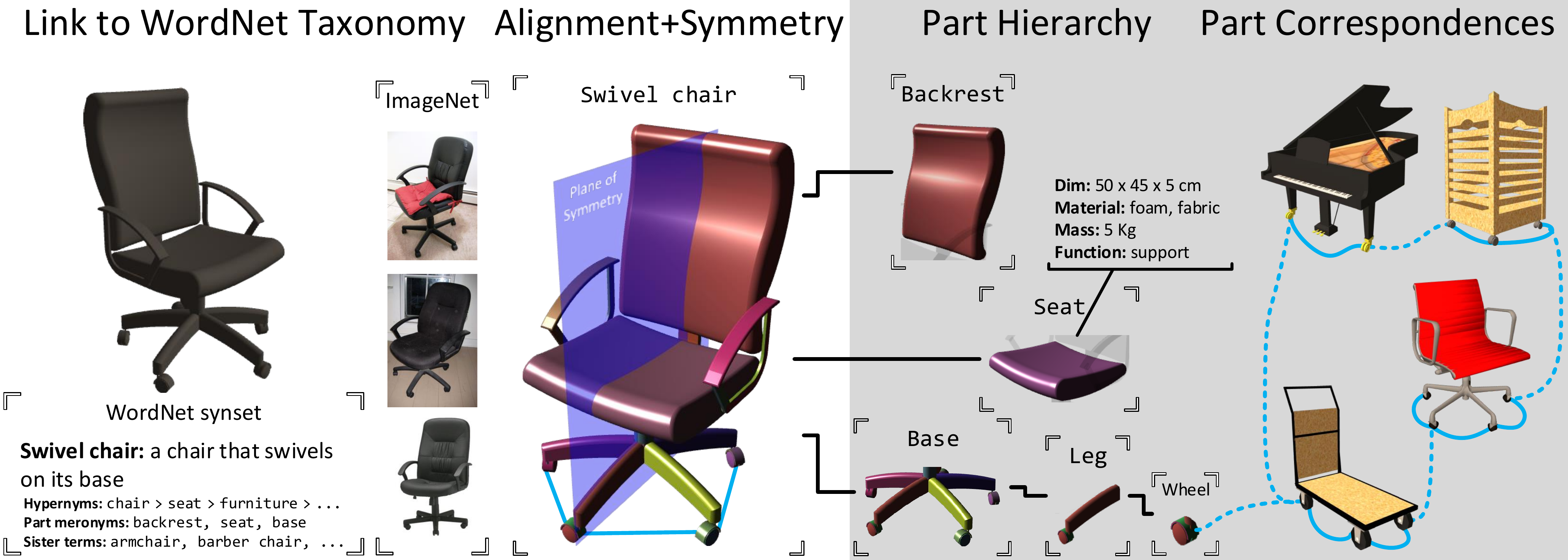}
  \caption{\small{ShapeNet annotations illustrated for an example chair model. \emph{Left:} links to the WordNet taxonomy provide definitions of objects, \texttt{is-a} and \texttt{has-a} relations, and a connection to images from ImageNet. \emph{Middle-left:} shape is aligned to a consistent upright and front orientation, and symmetries are computed \emph{Middle-right:} hierarchical decomposition of shape into parts on which various attributes are defined: names, symmetries, dimensions, materials, and masses. \emph{Right:} part-to-part and point-to-point connections are established at all levels within ShapeNet producing a dense and semantically rich network of correspondences. The gray background indicates annotations that are currently ongoing and not yet available for release.}}
  \label{fig:shapenet-view}
\end{figure*}

We envision ShapeNet as far more than a collection of 3D models. ShapeNet will include a rich set of annotations that provide semantic information about those models, establish links between them, and links to other modalities of data (e.g., images). These annotations are exactly what make ShapeNet uniquely valuable.  \Cref{fig:shapenet-view} illustrates the value of this dense network of interlinked attributes on shapes, which we describe below.

\mypara{Language-related Annotations:} Naming objects by their basic category is useful for indexing, grouping, and linking to related sources of data.  As described in the previous section, we organize ShapeNet based on the WordNet~\cite{miller:1995:wordnet} taxonomy.  Synsets are interlinked with various relations, such as hyper and hyponym, and part-whole relations.  Due to the popularity of WordNet, we can leverage other resources linked to WordNet such as ImageNet, ConceptNet, Freebase, and Wikipedia. In particular, linking to ImageNet~\cite{deng:2009:imagenet} will help transport information between images and shapes.  We assign each 3D model in ShapeNet to one or more synsets in the WordNet taxonomy (i.e., we populate each synset with a collection of shapes). Please refer to \Cref{sec:acquisition:categoryannotation} for details on the acquisition and validation of basic category annotations.  Future planned annotations include natural language descriptions of objects and object part-part relation descriptions.

\mypara{Geometric Annotations:} A critical property that distinguishes ShapeNet from image and video datasets is the fidelity with which 3D geometry represents real-world structures. We combine algorithmic predictions and manual annotations to organize shapes by category-level geometric properties and further derive rich geometric annotations from the raw 3D model geometry.
\begin{itemize}
\item \textbf{Rigid Alignments:}
Establishing a consistent canonical orientation (e.g., upright and front) for every model is important for various tasks such as visualizing shapes~\cite{Kim:2012:fuzzycorres}, shape classification~\cite{huang:2013:fineGrained} and shape recognition~\cite{wu:2014:3d}. Fortunately, most raw 3D model data is by default placed in an upright orientation, and the front orientations are typically aligned with an axis.  This allows us to use a hierarchical clustering and alignment approach to ensure consistent rigid alignments within each category (see \Cref{sec:alignment}).
\item \textbf{Parts and Keypoints:} Many shapes contain or have natural decompositions into important parts, as well as significant keypoints related to both their geometry and their semantics. For example, often different materials are associated with different parts. We intend to capture as much of that as possible into ShapeNet.
\item \textbf{Symmetry:} Bilateral symmetry planes and rotational symmetries are prevalent in artificial and natural objects, and deeply connected with the alignment and functionality of shapes. We refer to \Cref{sec:symmetry} for more details on how we compute symmetries for the shapes in ShapeNet.
\item \textbf{Object Size:} Object size is useful for many applications, such as reducing the hypothesis space in object recognition. Size annotations are discussed in \Cref{sec:shapenetsem}.
\end{itemize}

\mypara{Functional Annotations:} Many objects, especially man-made artifacts such as furniture and appliances, can be used by humans. Functional annotations describe these usage patterns. Such annotations are often highly correlated with specific regions of an object. In addition, it is often related with the specific type of human action. ShapeNet aims to store functional annotations at the global shape level and at the object part level.
\begin{itemize}
\item \textbf{Functional Parts:} Parts are critical for understanding object structure, human activities involving a 3D shape, and ergonomic product design. We plan to annotate parts according to their function --- in fact the very definition of parts has to be based on both geometric and functional criteria.
\item \textbf{Affordances:} We are interested in affordance annotations that are function and activity specific. Examples of such annotations include supporting plane annotations, and graspable region annotations for various object manipulations.
\end{itemize}

\mypara{Physical Annotations:} Real objects exist in the physical world and typically have fixed physical properties such as dimensions and densities.  Thus, it is important to store physical attribute annotations for 3D shapes.
\begin{itemize}
\item \textbf{Surface Material:} We are especially interested in the optical properties and semantic names of surface materials. They are important for applications such as rendering and structural strength estimation.
\item \textbf{Weight:} A basic property of objects which is very useful for physical simulations, and reasoning about stability and static support.
\end{itemize}

In general, the issue of compact and informative representations for all the above attributes over shapes raises many interesting questions that we will need to address as part of the ShapeNet effort.  Many annotations are currently ongoing projects and involve interesting open research problems.

\subsection{Annotation Methodology}

Though at first glance it might seem reasonable to collect the annotations we describe purely through manual human effort, we will in general take a hybrid approach.  For annotation types where it is possible, we will first algorithmically predict the annotation for each model instance (e.g., global symmetry planes, consistent rigid alignments).  We will then verify these predictions through crowd-sourcing pipelines and inspection by human experts.  This hybrid strategy is sensible in the context of 3D shape data as there are already various algorithms we can leverage, and collecting corresponding annotations entirely through manual effort can be extremely labor intensive.  In particular, since objects in a 3D representation are both more pure and more complete than objects in images, we can expect better and easier to establish correspondences between 3D shapes, enabling algorithmic transport of semantic annotations.  In many cases, the design of the human annotation interfaces themselves is an open question --- which stands in contrast to largely manual image labeling efforts such as ImageNet.  As a concrete example, shape part annotation can be presented and performed in various ways with different trade-offs in the type of obtained part annotation, the accuracy and the efficiency of the annotation process.

Coupled with this hybrid annotation strategy, we also take particular care to preserve the provenance and confidence of each algorithmic and human annotation.  The annotation source (whether an algorithm, or human effort), and a measure of the trust we can place in each annotation are critical pieces of information especially when we have to combine, aggregate, and reconcile several annotations.

\subsection{Annotation Schema and Web API}

To provide convenient access to all of the model and annotation data contained within ShapeNet, we construct an index over all the 3D models and their associated annotations using the Apache Solr framework.\footnote{\url{http://lucene.apache.org/solr/}}  Each stored annotation for a given 3D model is contained within the index as a separate attribute that can be easily queried and filtered through a simple web-based UI.  In addition, to make the dataset conveniently accessible to researchers, we provide a batched download capability.

\section{Annotation Acquisition and Validation}
\label{sec:acquisition}
A key challenge in constructing ShapeNet is the methodology for acquiring and validating annotations. Our goal is to provide all annotations with high accuracy. In cases where full verification is not yet available, we aim to estimate a confidence metric for each annotation, as well as record its provenance. This will enable others to properly estimate the trustworthiness of the information we provide and use it for different applications.

\subsection{Category Annotation}
\label{sec:acquisition:categoryannotation}
As described in Section~\ref{sec:annotation-types}, we assign each 3D model to one or more synsets in the WordNet taxonomy.

\mypara{Annotation}
Models are retrieved by textual query into the online repositories that we collected, and the initial category annotation is set to the used textual query for each retrieved model.  After we retrieve these models we use the popularity score of each model on the repository to sort models and ask human workers to verify the assigned category annotation.  This is sensible since the more popular models tend to be high quality and correctly retrieved through the category keyword textual query.  We stop verifying category annotations with people once the positive ratio is lower than a 2\% threshold.

\mypara{Clean-up}
In order for the dataset to be easily usable by researchers it should contain clean and high quality 3D models.  Through inspection, we identify and group 3D models into the following categories: \emph{single 3D models}, \emph{3D scenes}, \emph{billboards}, and \emph{big ground plane}.

\begin{itemize}\denselist
  \item Single 3D models: semantically distinct objects; focus of our \emph{ShapeNetCore} annotation effort.
  \item 3D scenes: detected by counting the number of connected components in a voxelized representation. We manually verify these detections and mark scenes for future analysis.
  \item Billboards: planes with a painted texture. Often used to represent people and trees. These models are generally not useful for geometric analysis. They can be detected by checking whether a single plane can fit all vertices.
  \item Big ground plane: object of interest placed on a large horizontal plane or in front of large vertical plane. Although we do not currently use these models, the plane can easily be identified and removed through simple geometric analysis.
\end{itemize}

We currently include the single 3D models in the ShapeNetCore subset of ShapeNet.

\subsection{Hierarchical Rigid Alignment}
\label{sec:alignment}
\begin{figure}[t]
	\includegraphics[width=\linewidth]{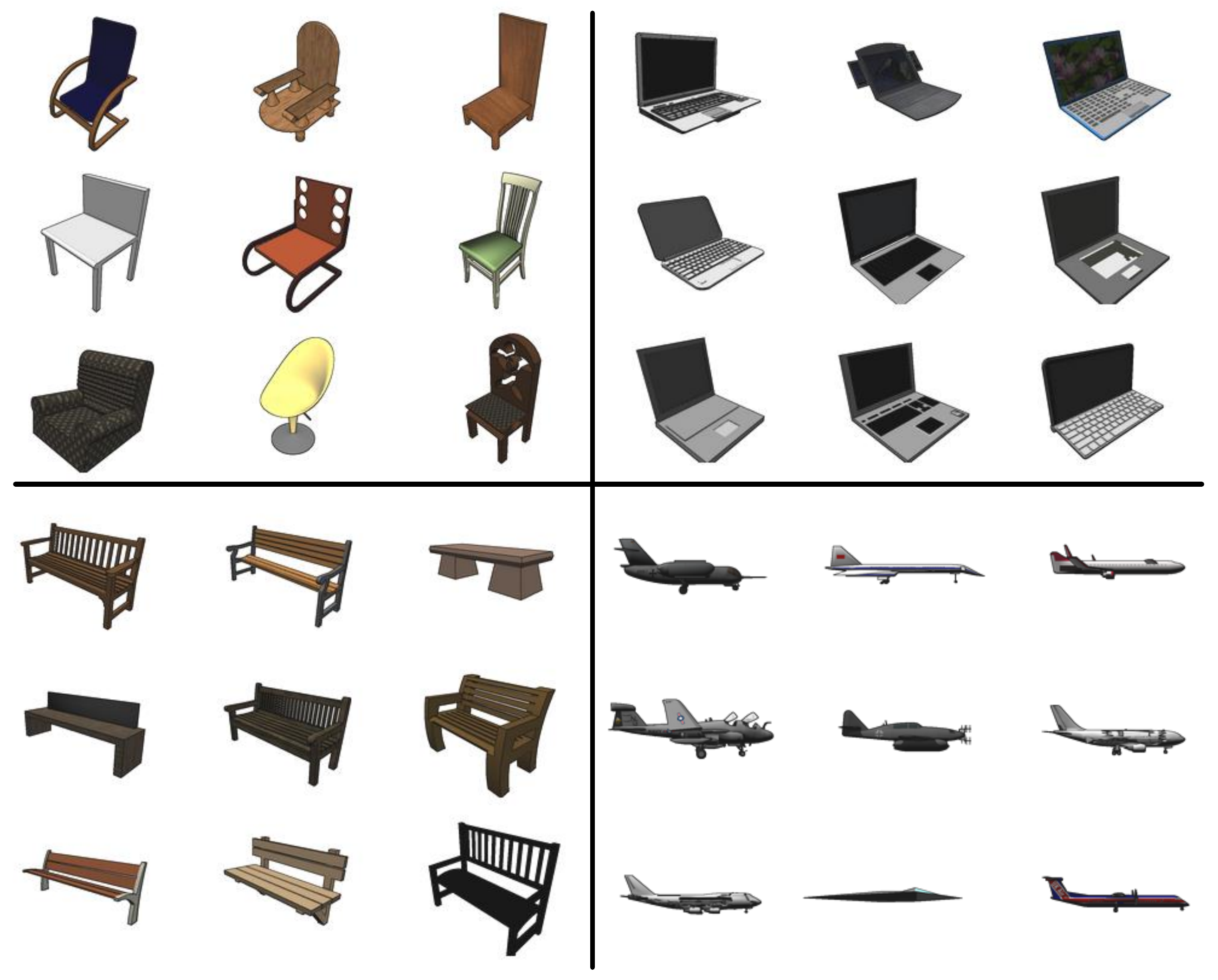}
	\caption{Examples of aligned models in the chair, laptop, bench, and airplane synsets.}
	\label{fig:rigid_alignment}
\end{figure}

The goal of this step is to establish a consistent canonical orientation for models within each category. Such alignment is important for various tasks such as visualizing shapes, shape classification and shape recognition. \Cref{fig:rigid_alignment} shows several categories in ShapeNet that have been consistently aligned.

Though the concept of consistent orientation seems natural, one issue has to be addressed. We explain by an example. ``armchair'', ``chair'' and ``seat'' are three categories in our taxonomy, each being a subcategory of its successor. Consistent orientation can be well defined for shapes in the ``armchair'' category, by checking arms, legs and backs. Yet, it becomes difficult to define for the ``chair'' category. For example, ``side chair'' and ``swivel chair'' are both subcategories of ``chair'', however, swivel chairs have a very different leg structure than most side chairs. It becomes even more ambiguous to define for ``seat'', which has subcategories such as ``stool'', ``couch'', and ``chair''. However, the concept of an upright orientation still applies throughout most levels of the taxonomy.

Following the above discussion, it is natural for us to propose a hierarchical alignment method, with a small amount of human supervision.  The basic idea is to hierarchically align models following the taxonomy of ShapeNet in a bottom-up manner, i.e., we start from aligning shapes in low-level categories and then gradually elevate to higher level categories. When we proceed to the higher level, the self-consistent orientation within a subcategory should be maintained. For the alignment at each level, we first use a geometric algorithm described in the \Cref{sec:appendix:alignment}, and then ask human experts to check and correct possible misalignments.  With this strategy, we efficiently obtain consistent orientations.  In practice, most shapes in the same low-level categories can be well aligned algorithmically, requiring limited manual correction.  Though the proportion of manual corrections increases for aligning higher-level categories, the number of categories at each level becomes logarithmically smaller.

\subsection{Parts and Keypoints}
To obtain part and keypoint annotations we start from some curated part annotations within each category. For parts, this acquisition can be speeded up by having  algorithmically generated segmentations and then having users accept or modify parts from these. We intend to experiment with both 2D and 3D interfaces for this task. We then exploit a number of different algorithmic techniques to propagate this information to other nearby shapes. Such methods can rely on rigid alignments in 3D, feature descriptor alignments in an appropriately defined feature space, or general shape correspondences. We iterate this pipeline, using active learning to estimate the 3D models and regions of these models where further human annotation would be most informative, generate a new set of crowd-sourced annotation tasks, algorithmically propagate their results, and so on. In the end we have users verify all proposed parts and keypoints, as verification is much faster than direct annotation.
\label{sec:parts}

\subsection{Symmetry Estimation}
We provide bilateral symmetry plane detections for all 3D models in ShapeNetCore. Our method is a modified version of~\cite{mitra:2013:symmetry}. The basic idea is to use hough transform to vote on the parameters of the symmetry plane. More specifically, we generate all combinations of pairs of vertices from the mesh. Each pair casts a vote of a possible symmetry plane in the discretized space of plane parameters partitioned evenly. We then pick the parameter with the most votes as the symmetry plane candidate. As a final step, this candidate is verified to ensure that every vertex has a symmetric counterpart.
\label{sec:symmetry}

\subsection{Physical Property Estimation}
Before computing physical attribute annotations, the dimensions of the models need to be correspond to the real world.  We estimate the absolute dimensions of models using prior work in size estimation~\cite{savva:2014:sizes}, followed by manual verification.  With the given absolute dimensions, we now compute the total solid volume of each model through filled-in voxelization.  We use the space carving approach implemented by Binvox~\cite{nooruddin:2003:simplification}.  Categories of objects that are known to be container-like (i.e., bottles, microwaves) are annotated as such and only the surface voxelization volume is used instead.  We then estimate the proportional material composition of each object category and use a table of material densities along with each model instance volume to compute a rough total weight estimate for that instance.  More details about the acquisition of these physical attribute annotations are available separately~\cite{savva:2015:semgeo}. 
\label{sec:physical}



\section{Current Statistics}
\label{sec:statistics}
At the time of this technical report, ShapeNet has indexed roughly 3,000,000 models. 220,000 models of these models are classified into 3,135 categories (WordNet synsets). Below we provide detailed statistics for the currently annotated models in ShapeNet as a whole, as well as details of the available publicly released subsets of ShapeNet.

\begin{figure*}[t]
	\includegraphics[width=\linewidth]{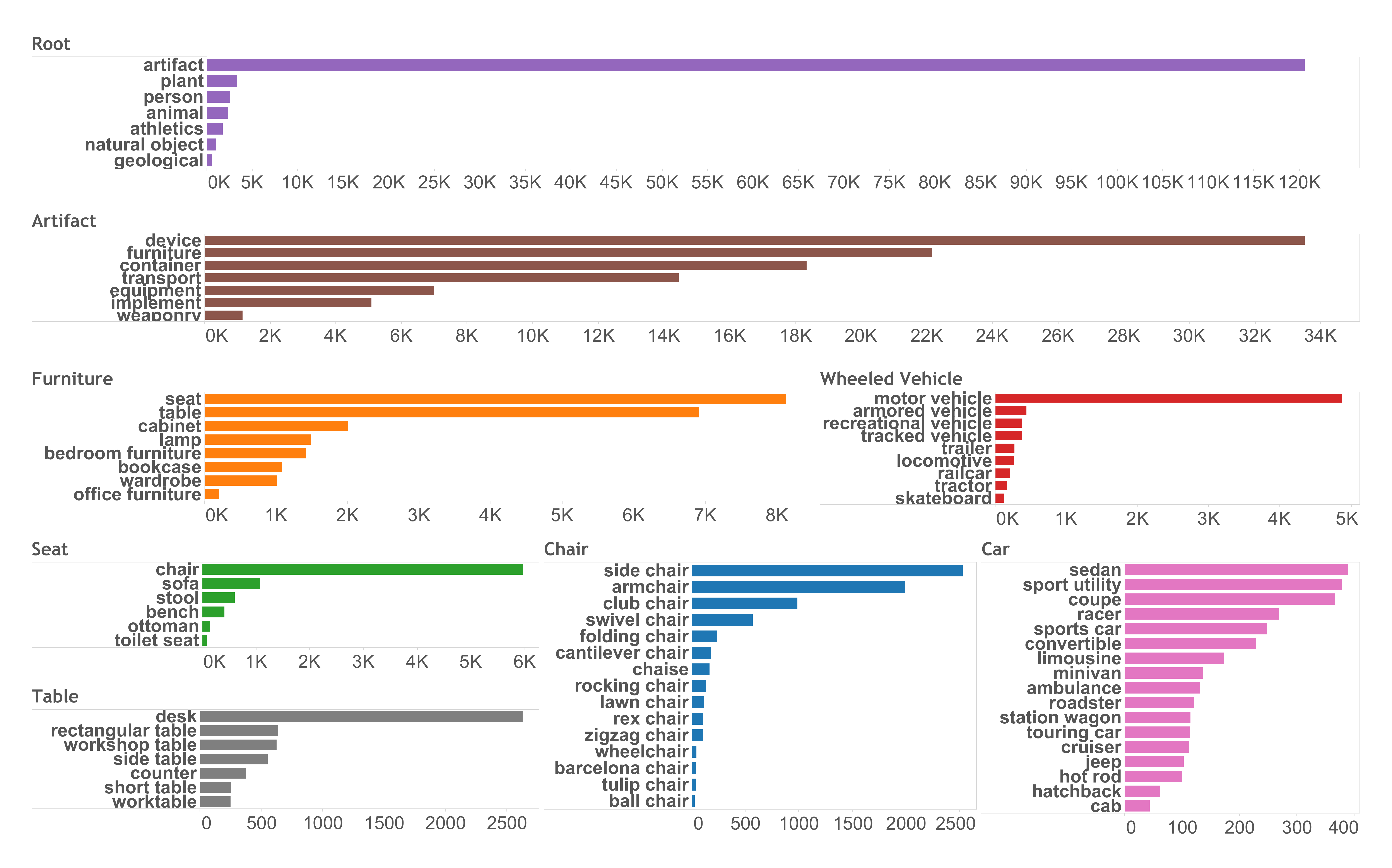}
	\caption{Plots of the distribution of ShapeNet models over WordNet synsets at multiple levels of the taxonomy (only the top few children synsets are shown at each level). The highest level (root) is at the top and the taxonomy levels become lower downwards and to the right.  Note the bias towards rigid man-made artifacts at the top and the broad coverage of many low level categories towards the bottom.}
	\label{fig:category-distributions}
\end{figure*}

\mypara{Category Distribution} \Cref{fig:category-distributions} shows the distributions of the number of shapes per synset at various taxonomy levels for the current ShapeNetCore corpus. To the best of our knowledge, ShapeNet is the largest clean shape dataset available in terms of total number of shapes, average number of shapes per category, as well as the number of categories.

We observe that ShapeNet as a whole is strongly biased towards categories of rigid man-made artifacts, due to the bias of the source 3D model repositories.  This is in contrast to common image database statistics that contain more natural objects such as plants and animals~\cite{torralba:2010:labelme}. This distribution bias is probably due to a combination of factors: 1) meshes of natural objects are more difficult to design using common CAD software; 2) 3D model consumers are typically more interested in artificial objects such as those observed in modern urban lifestyles. The former factor can be mitigated in the near future by using the rapidly improving depth sensing and 3D scanning technology.

\subsection{ShapeNetCore}

\emph{ShapeNetCore} is a subset of the full ShapeNet dataset with single clean 3D models and manually verified category and alignment annotations. It covers 55 common object categories with about 51,300 unique 3D models. The 12 object categories of PASCAL 3D+\cite{xiang:2014:pascal3d}, a popular computer vision 3D benchmark dataset, are all covered by ShapeNetCore.  The category distribution of ShapeNetCore is shown in \Cref{tab:shapenetcore}.

\begin{table*}
\centering{
	\begin{tabular}{ccr|ccr|ccr}
	\hline
	ID&Name&Num&ID&Name&Num&ID&Name&Num\\
	\hline
	04379243&table&8443	&	03593526&jar&597	&04225987&skateboard&152\\
	02958343&car&7497	&	02876657&bottle&498	&04460130&tower&133\\
	03001627&chair&6778	&	02871439&bookshelf&466	&02942699&camera&113\\
	02691156&airplane&4045	&	03642806&laptop&460	&02801938&basket&113\\
	04256520&sofa&3173	&	03624134&knife&424	&02946921&can&108\\
	04090263&rifle&2373	&	04468005&train&389	&03938244&pillow&96\\
	03636649&lamp&2318	&	02747177&trash bin&343	&03710193&mailbox&94\\
	04530566&watercraft&1939	&	03790512&motorbike&337	&03207941&dishwasher&93\\
	02828884&bench&1816	&	03948459&pistol&307	&04099429&rocket&85\\
	03691459&loudspeaker&1618	&	03337140&file cabinet&298	&02773838&bag&83\\
	02933112&cabinet&1572	&	02818832&bed&254	&02843684&birdhouse&73\\
	03211117&display&1095	&	03928116&piano&239	&03261776&earphone&73\\
	04401088&telephone&1052	&	04330267&stove&218	&03759954&microphone&67\\
	02924116&bus&939	&	03797390&mug&214	&04074963&remote&67\\
	02808440&bathtub&857	&	02880940&bowl&186	&03085013&keyboard&65\\
	03467517&guitar&797	&	04554684&washer&169	&02834778&bicycle&59\\
	03325088&faucet&744	&	04004475&printer&166	&02954340&cap&56\\	
	03046257&clock&655	&	03513137&helmet&162	& & &\\
	03991062&flowerpot&602	&	03761084&microwaves&152 & &\bf{Total}&57386 \\
	\hline\\
	\end{tabular}
	\caption{\small{Statistics of ShapeNetCore synsets. ID corresponds to WordNet synset offset, which is aligned with ImageNet.}}
	\label{tab:shapenetcore}
}
\end{table*}

\subsection{ShapeNetSem}
\label{sec:shapenetsem}

\begin{table*}
\centering\footnotesize
\begin{tabular}{cr|cr|cr|cr|cr}
	\hline
	\textbf{Category}&\textbf{Num}&\textbf{Category}&\textbf{Num}&\textbf{Category}&\textbf{Num}&\textbf{Category}&\textbf{Num}&\textbf{Category}&\textbf{Num}\\
Chair&696&Monitor&127&WallLamp&78&Gun&54&FlagPole&38\\
Lamp&663&RoundTable&120&SideChair&77&Nightstand&53&TvStand&38\\
ChestOfDrawers&511&TrashBin&117&VideoGameConsole&75&Mug&51&Fireplace&37\\
Table&427&DrinkingUtensil&112&MediaStorage&73&AccentChair&50&Rack&37\\
Couch&413&DeskLamp&110&Painting&73&ChessBoard&49&LightSwitch&36\\
Computer&244&Clock&101&Desktop&71&Rug&49&Oven&36\\
Dresser&234&ToyFigure&101&AccentTable&70&WallUnit&46&Airplane&35\\
TV&233&Plant&98&Camera&70&Mirror&45&DresserWithMirror&35\\
WallArt&222&Armoire&95&Picture&69&Bowl&44&Calculator&34\\
Bed&221&QueenBed&94&Refrigerator&68&SodaCan&44&TableClock&34\\
Cabinet&221&Stool&92&Speaker&68&VideoGameController&44&Toilet&34\\
FloorLamp&201&EndTable&91&Sideboard&67&WallClock&43&Cup&33\\
Desk&189&Bottle&88&Barstool&66&Printer&42&Stapler&33\\
PottedPlant&188&DiningTable&88&Guitar&65&Sword&40&PaperBox&32\\
FoodItem&180&Bookcase&87&MediaPlayer&62&USBStick&40&SpaceShip&32\\
Laptop&173&CeilingLamp&86&Ipod&59&Chaise&39&Toy&32\\
Vase&163&Bench&84&PersonStanding&57&OfficeSideChair&39&ToiletPaper&31\\
TableLamp&142&Book&84&Piano&56&Poster&39&Knife&30\\
OfficeChair&137&CoffeeTable&81&Curtain&55&Sink&39&PictureFrame&30\\
CellPhone&130&Pencil&80&Candle&54&Telephone&39&Recliner&30\\

	\hline\\
\end{tabular}
\caption{\small{Total number of models for the top 100 ShapeNetSem categories (out of 270 categories). Each category is also linked to the corresponding WordNet synset, establishing the same linkage to WordNet and ImageNet as with ShapeNetCore.}}
\label{tab:shapenetsem}
\end{table*}

\emph{ShapeNetSem} is a smaller, more densely annotated subset consisting of 12,000 models spread over a broader set of 270 categories.  In addition to manually verified category labels and consistent alignments, these models are annotated with real-world dimensions, estimates of their material composition at the category level, and estimates of their total volume and weight.  The total numbers of models for the top 100 categories in this subset are given in \Cref{tab:shapenetsem}.


\section{Discussion and Future Work}
\label{sec:discussion}
The construction of ShapeNet is a continuous, ongoing effort.  Here we have just described the initial steps we have taken in defining ShapeNet and populating a core subset of model annotations that we hope will prove useful to the community.  We plan to grow ShapeNet in four distinct directions:

\mypara{Additional annotation types} We will introduce several additional types of annotations that have strong connections to the semantics and functionality of objects. Firstly, hierarchical part decompositions of objects will provide a useful finer granularity description of object structure that can be leveraged for part segmentation and shape synthesis. Secondly, physical object property annotations such as materials and their attributes will allow higher fidelity physics and appearance simulation, adding another layer of understanding to methods in vision and graphics.

\mypara{Correspondences} One of the most important goals of ShapeNet is to provide a dense network of correspondences between 3D models and their parts.  This will be invaluable for enabling much shape analysis research and helping to improve and evaluate methods for many traditional tasks such as alignment and segmentation.  Additionally, we plan to provide correspondences between 3D model parts and image patches in ImageNet --- a link that will be critical for propagating information between image space and 3D models.

\mypara{RGB-D data} The rapid proliferation of commodity \mbox{RGB-D} sensors is already making the process of capturing real-world environments better and more efficient.  Expanding ShapeNet to include shapes reconstructed from scanned \mbox{RGB-D} data is a critical goal.  We foresee that over time, the amount of available reconstructed shape data will overshadow the existing designed 3D model data and as such this is a natural growth direction for ShapeNet.  A related effort that we are currently undertaking is to align 3D models to objects observed in RGB-D frames.  This will establish a powerful connection between real world observations and 3D models.

\mypara{Annotation coverage} We will continue to expand the set of annotated models to cover a bigger subset of the entirety of ShapeNet. We will explore combinations of algorithmic propagation methods and crowd-sourcing for verification of the algorithmic results.

\vfill

\section{Conclusion}
\label{sec:conclusion}
We firmly believe that ShapeNet will prove to be an immensely useful resource to several research communities in several ways:

\mypara{Data-driven research} By establishing ShapeNet as the first large-scale 3D shape dataset of its kind we can help to move computer graphics research toward a data-driven direction following recent developments in vision and NLP. Additionally, we can help to enable larger-scale quantitative analysis of proposed systems that can clarify the benefits of particular methodologies against a broader and more representative variety of 3D model data.

\mypara{Training resource} By providing a large-scale, richly annotated dataset we can also promote a broad class of recently resurgent machine learning and neural network methods for applications dealing with geometric data. Much like research in computer vision and natural language understanding, computational geometry and graphics stand to benefit immensely from these data-driven learning approaches.

\mypara{Benchmark dataset} We hope that ShapeNet will grow to become a canonical benchmark dataset for several evaluation tasks and challenges.  In this way, we would like to engage the broader research community in helping us define and grow ShapeNet to be a pivotal dataset with long-lasting impact.

\bibliographystyle{plain}
\begin{small}
\bibliography{main}
\end{small}

\newpage
\appendix
\section{Appendix}
\subsection{Hierarchical Rigid Alignment}
\label{sec:appendix:alignment}

In the following, we describe our hierarchical rigid alignment algorithm in more detail.

As a pre-processing step, we first semi-automatically align the upright orientation of each shape. Fortunately, most shapes downloaded from the web are by default placed in the upright orientations. For those that are not, we filter them out by manual inspection. We then convert models to point clouds through furthest point sampling and perform PCA on the point sets. Finally, we ask a person to pick the vector of correct upright orientation from six candidates containing the PCA axes and their reverse directions.

Starting from a leaf category in ShapeNet, we jointly align all shapes following prior work~\cite{huang:2013:fineGrained}. If a leaf category has more than 100 shapes, we further partition it into smaller, more coherent clusters by $k$-means clustering using pose-invariant global features, such as phase-invariant HoG features [see appendix]. Here we briefly review~\cite{huang:2013:fineGrained}. Each shape is associated with a random variable, denoting the transformation of the shape from its original pose to the consistent canonical pose. Over the set of shapes, a Markov Random Field (MRF) is constructed, whose energy function measures the consistency of all pairs of shapes after applying their transformations. In practice, the space of rigid transformations is discretized into $N$ bins. We perform MAP inference over the MRF to find the optimal transformation for each shape. We then manual inspect the results and correct occasional errors.

After this step, we represent each leaf node category by the shape in the centroid of the feature space. Then, we gather the representative shapes for all leaf categories of an intermediate category and apply~\cite{huang:2013:fineGrained} again for joint alignment. This higher-level algorithmic alignment is verified by a person again. The procedure is applied along the taxonomy hierarchy until the root node is reached.

\end{document}